\documentstyle[12pt,fleqn,epsf,epsfig,axodraw]{article}
\def\lsim{\raise0.3ex\hbox{$\;<$\kern-0.75em\raise-1.1ex\hbox{$\sim\;$}}}
\def\gsim{\raise0.3ex\hbox{$\;>$\kern-0.75em\raise-1.1ex\hbox{$\sim\;$}}}

\begin{document}
\setlength{\unitlength}{1cm}
\setlength{\mathindent}{0cm}
\thispagestyle{empty}

\begin{flushright}
	UWThPh-2004-15\\
	WUE-ITP-04-016\\
	HEPHY-PUB 791/04\\
	hep-ph/0406309
\end{flushright}


\begin{center}
	{\Large \bf 
		CP violation in chargino production \\
		and decay into sneutrino
		}
\vskip 2.0em
{\large
{\sc A.~Bartl$^{a}$\footnote{e-mail:
        bartl@ap.univie.ac.at}
     H.~Fraas$^{b}$\footnote{e-mail:
        fraas@physik.uni-wuerzburg.de},
	  O.~Kittel$^{b}$\footnote{e-mail:
		  kittel@physik.uni-wuerzburg.de},
	  W.~Majerotto$^{c}$\footnote{e-mail:
        majer@qhepu3.oeaw.ac.at}
}}\\[1ex]
{\normalsize \it
$^{a}$ Institut f\"ur Theoretische Physik, Universit\"at Wien, 
Boltzmanngasse 5, A-1090 Wien, Austria}\\
{\normalsize \it
$^{b}$ Institut f\"ur Theoretische Physik, Universit\"at
W\"urzburg, Am Hubland, D-97074~W\"urzburg, Germany}\\
{\normalsize \it
$^{c}$ Institut f\"ur Hochenergiephysik, \"Osterreichische
Akademie der Wissenschaften, Nikolsdorfergasse 18, 
A-1050 Wien, Austria}\\
\end{center}
\vskip 2.0em

\begin{abstract}

We study CP odd asymmetries  in chargino production 
$e^+e^- \to\tilde\chi^{\pm}_1 \tilde\chi^{\mp}_2$
and the subsequent two-body decay of one chargino
into a sneutrino. 
We show that in the Minimal Supersymmetric Standard Model with complex  
parameter $\mu$ the asymmetries can reach  $30\%$. 
We discuss the feasibility of measuring these asymmetries
at a  linear collider with $\sqrt{s}=800$ GeV and 
longitudinally polarized beams.

\end{abstract}


\section{Introduction}

In the the chargino sector of the
Minimal Supersymmetric Standard Model (MSSM) \cite{haberkane}
the Higgsino mass parameter $\mu$ can be complex \cite{Dugan}.
It has been shown  that in the production  of two different charginos,
$e^+e^-\to\tilde\chi^{\pm}_1\tilde\chi^{\mp}_2$, a CP violating phase  
$\varphi_{\mu}$ of $\mu$ causes a non-vanishing chargino polarization 
perpendicular to the production plane
\cite{choi1,choigaiss}.
This polarization  leads at tree level to triple product asymmetries 
\cite{tripleprods,karl,olaf1},
which might be large and will allow us to constrain $\varphi_{\mu}$ 
at a future $e^+e^-$ linear collider \cite{LC}.
Usually it is claimed that this phase has to be small for a light 
supersymmetric (SUSY) particle spectrum due to the experimental upper 
bounds of the electric dipole moments (EDMs) \cite{edms}. However, these
restrictions are model dependent \cite{nath}. If cancellations among 
different contributions occur and, for example, if lepton
flavor violating phases are present, the EDM restrictions
on $\varphi_{\mu}$ may disappear \cite{BMPW}.

We study chargino production
\begin{eqnarray} \label{production}
	e^++e^-&\to&\tilde\chi^+_i+\tilde\chi^-_j, 
	\quad i,j =1,2,  
\end{eqnarray}
with longitudinally polarized beams and
the subsequent  two-body decay of one of the 
charginos into a sneutrino
\begin{eqnarray} \label{decay_1A}
	\tilde\chi^+_i \to \ell^+ + \tilde\nu_{\ell}, 
	\quad \ell = e,\mu,\tau.
\end{eqnarray}
We define the triple product
 \begin{eqnarray}\label{tripleproduct1}
	 {\mathcal T}_{\ell} &=& 
	 (\vec p_{e^-} \times \vec p_{\tilde\chi^+_i}) \cdot \vec p_{\ell}
 \end{eqnarray}
and the T odd asymmetry 
\begin{eqnarray}\label{AT1}
	 {\mathcal A}_{\ell}^{\rm T} &=& 
 \frac{\sigma({\mathcal T}_{\ell}>0)-\sigma({\mathcal T}_{\ell}<0)}
      {\sigma({\mathcal T}_{\ell}>0)+\sigma({\mathcal T}_{\ell}<0)},
\end{eqnarray}
of the cross section  $\sigma$ for chargino production
(\ref{production}) and decay (\ref{decay_1A}).
The asymmetry ${\mathcal A}_{\ell}^{\rm T}$ is not only sensitive
to the phase  $\varphi_{\mu}$, but also to absorptive contributions,
which could enter via s-channel resonances or final-state interactions.
In order to eliminate the contributions from  
the absorptive parts, which do not signal CP violation,
we will study the CP asymmetry
\begin{equation}
{\mathcal A}_{\ell} = 
\frac{1}{2}({\mathcal A}_{\ell}^{\rm T}-\bar{\mathcal A}_{\ell}^{\rm T}),
\label{ACP1}
\end{equation}
where $\bar{\mathcal A}_{\ell}^{\rm T}$ is the asymmetry for
the CP conjugated process 
$e^+e^-\to\tilde\chi^-_i\tilde\chi^+_j; \; 
\tilde\chi^-_i \to  \ell^-\, \bar{\tilde\nu_{\ell}}.$ 
In this context it is interesting to note that in chargino production it
is not possible to construct a triple product and a corresponding
asymmetry by using transversely polarized $e^+$ and $e^-$ beams 
\cite{choi1,holger}, therefore, one has to rely on the transverse
polarization of the produced chargino.

In Section 
\ref{Definitions and Formalism} 
we give our definitions and formalism used,
and  the analytical  formulae for the 
chargino production and decay cross sections.
In Section 
3 we discuss some general
properties of the CP asymmetries. 
In Section \ref{Numerical results} we present numerical results  
for ${\mathcal A}_{\ell}$ and the cross sections. 
Section \ref{Summary and conclusion} gives a summary 
and conclusions.

\section{Definitions and formalism
  \label{Definitions and Formalism}}

\subsection{Lagrangians and couplings
     \label{Lagrangian}}

The MSSM interaction Lagrangians relevant for our study are 
(in our notation and conventions we follow 
closely \cite{haberkane,gudi}):
\begin{eqnarray}
{\cal L}_{Z^0 \ell \bar \ell} &=&
- \frac{g}{\cos\theta_W} Z_{\mu}\bar \ell\gamma^{\mu}[L_{\ell}P_L+
	R_{\ell} P_R]\ell,\\
{\cal L}_{Z^0\tilde\chi_j^{+}\tilde\chi_i^-} &=&
 \frac{g}{\cos\theta_W}Z_{\mu}\bar{\tilde\chi}^+_i\gamma^{\mu}
[O_{ij}^{'L} P_L+O_{ij}^{'R} P_R]\tilde\chi_j^{+},\\
{\cal L}_{\ell \tilde\nu_{\ell}\tilde\chi^+_i} &=&
- g V_{i1}^{*} \bar{\tilde\chi}_i^{+C} P_L \ell 
\tilde{\nu}^{*}_{\ell}+\mbox{h.c.}, \quad \ell=e,\mu, \\
{\cal L}_{\tau \tilde\nu_{\tau}\tilde\chi^+_i} &=&
- g \bar{\tilde\chi}_i^{+C}(V_{i1}^{*} P_L-Y_{\tau}U_{i2} P_R) \tau 
\tilde{\nu}^{*}_{\tau} +\mbox{h.c.},
\end{eqnarray}
with the couplings:
\begin{eqnarray}
 L_{\ell}&=&T_{3\ell}-e_{\ell}\sin^2\theta_W, \quad
 R_{\ell}\;=\;-e_{\ell}\sin^2\theta_W,\\
 O_{ij}^{'L}&=&-V_{i1} V_{j1}^{*}-\frac{1}{2} V_{i2} V_{j2}^{*}+
\delta_{ij} \sin^2\theta_W,\\
 O_{ij}^{'R}&=&-U_{i1}^{*} U_{j1}-\frac{1}{2} U_{i2}^{*} U_{j2}+
\delta_{ij} \sin^2\theta_W,
\end{eqnarray}
with $i,j=1,2$. Here 
$P_{L, R}=\frac{1}{2}(1\mp \gamma_5)$, $g=e/\sin\theta_W$ is 
the weak coupling constant, and $e_\ell$ and $T_{3 \ell}$ denote the
charge and the third component of the weak isospin of the 
lepton $\ell$. The $\tau$-Yukawa coupling is given by  
$Y_{\tau}= m_{\tau}/(\sqrt{2}m_W\cos\beta)$ with
$\tan\beta=\frac{v_2}{v_1}$, 
where $v_{1,2}$ are the vacuum expectation values of the two 
neutral Higgs fields. The chargino mass eigenstates 
$\tilde\chi^+_i={\chi_i^{+} \choose \bar\chi_i^{-}}$
are defined by $\chi^{+}_i=V_{i1}w^{+}+V_{i2} h^{+}$ and 
$\chi_j^{-}=U_{j1}w^{-}+U_{j2} h^{-}$ with $w^{\pm}$ and $h^{\pm}$
the two-component spinor fields of the W-ino and the charged
Higgsinos, respectively. The complex unitary $2\times 2$ matrices 
$U_{mn}$ and $V_{mn}$ diagonalize the chargino mass 
matrix $X_{\alpha\beta}$, $U_{m \alpha}^* X_{\alpha\beta}V_{\beta
n}^{-1}= m_{\tilde{\chi}^+_i}\delta_{mn}$, 
with $ m_{\tilde{\chi}^+_i}>0$. 

\subsection{Cross section
     \label{Cross section}}

We choose a coordinate frame such that in the laboratory system
the four momenta are:
   \begin{eqnarray}
  && p_{e^-}^{\mu} = E_b(1,-\sin\theta,0, \cos\theta),\quad
     p_{e^+}^{\mu} = E_b(1, \sin\theta,0,-\cos\theta),\\
 &&  p_{\tilde\chi^+_i}^{\mu} = (E_{\tilde\chi^+_i},0,0,-q),\quad
     p_{\tilde\chi^-_j}^{\mu} = (E_{\tilde\chi^-_j},0,0, q),
   \end{eqnarray}
with the beam energy $E_b=\sqrt{s}/2$, the scattering angle 
$\theta \angle (\vec p_{e^-},\vec p_{\tilde\chi^-_j})$ and 
the azimuth $\phi$ is chosen zero.
For the description of the polarization of chargino 
$\tilde\chi^+_i$ we choose three spin vectors in the laboratory system:
\begin{eqnarray}
	&&  s^{1,\mu}_{\tilde\chi^+_i}=(0,-1,0,0),\quad
    s^{2,\mu}_{\tilde\chi^+_i}=(0,0,1,0),\quad
	 s^{3,\mu}_{\tilde\chi^+_i}=
	 \frac{1}{m_{\tilde\chi^+_i}}(q,0,0,-E_{\tilde\chi^+_i}).
	 \label{spinvec}
\end{eqnarray} 
Together with  
$p_{\tilde\chi^+_i}^{\mu}/m_{\tilde\chi^+_i}$ they form an orthonormal set.

For the calculation of the cross section for the
combined process of chargino production (\ref{production})
and the subsequent two-body decay of
$\tilde\chi^+_i$ (\ref{decay_1A}),
we use the spin-density matrix formalism as in \cite{gudi,spin}.
The amplitude squared,  
\begin{eqnarray}       \label{amplitude}
|T|^2 &=&
|\Delta(\tilde\chi^+_i)|^2~
\sum_{\lambda_i,\lambda'_i}~
\rho_P   (\tilde\chi^+_i)^{\lambda_i \lambda_i'}\;
\rho_{D}(\tilde\chi^+_i)_{\lambda_i'\lambda_i},
\end{eqnarray}
is composed of the (unnormalized) spin-density production matrix
$\rho_P(\tilde\chi^+_i)$ and the decay matrix
$\rho_{D}(\tilde\chi^+_i)$, with the helicity indices 
$\lambda_i$ and $ \lambda_i'$ of the chargino.
The propagator is given by
$ \Delta(\tilde{\chi}^+_i)=i/[p_{\tilde\chi^+_i}^2-m_{\tilde\chi^+_i}^2
	+im_{\tilde\chi^+_i}\Gamma_{\tilde\chi^+_i}]$.
The production matrix 
$\rho_P(\tilde\chi^+_i)$ can be expanded in terms of the Pauli matrices
$\sigma^{a}$, $a=1,2,3$:
\begin{eqnarray} \label{rhoP}
  \rho_P(\tilde\chi^+_i)^{\lambda_i \lambda_i'} &=&
  2(\delta_{\lambda_i \lambda_i'} P + \sum_a
       \sigma^{a}_{\lambda_i \lambda_i'}
		\Sigma_P^a).   
\end{eqnarray}
With our choice of the spin vectors 
$s^a_{\tilde\chi^+_i}$, Eq.~(\ref{spinvec}),
$\Sigma^{3}_P/P$ is the longitudinal polarization of $ \tilde \chi^+_i$
in the laboratory system,
$\Sigma^{1}_P/P$ is the transverse polarization in the 
production plane and $\Sigma^{2}_P/P$ is the polarization
perpendicular to the production plane.
The analytical formulae for the expansion 
coefficients $P$ and $\Sigma^{a}_P$ are given in  \cite{gudi}.
The coefficient $\Sigma^{2}_P$ is non-zero only for 
production of an unequal pair of charginos,
$e^+e^- \to\tilde\chi^{\pm}_1 \tilde\chi^{\mp}_2$,
and obtains contributions from $Z$-exchange and $Z$-$\tilde \nu$ 
interference only \cite{gudi}:
\begin{eqnarray}
     \Sigma_P^2 &=&\Sigma_P^2(Z Z)+ \Sigma_P^2(Z\tilde \nu),
\end{eqnarray}
with
\begin{eqnarray}
\Sigma_P^2(Z Z)&=&2\frac{g^4}{\cos^4\theta_W}|\Delta (Z)|^2
	(c_{R}^{ZZ} - c_{L}^{ZZ})
	Im\Big\{O^{'L}_{ij}O^{'R\ast}_{ij}\Big\}
	E_b^2m_{\tilde\chi^-_j}q\sin\theta, \label{Zexchange} \\
\Sigma_P^2(Z\tilde \nu)&=&\frac{g^4}{\cos^2\theta_W}c_{L}^{Z\tilde \nu} 
	Im\Big\{V^{\ast}_{i1}V_{j1}O^{'R}_{ij}
	\Delta (Z)\Delta (\tilde \nu)^{\ast}\Big\}
	E_b^2m_{\tilde\chi^-_j}q\sin\theta. \label{ZNuinterference}
\end{eqnarray}
The propagators are defined by
    \begin{equation}
		      \Delta(Z)  = \frac{i}{p_Z^2-m^2_Z+im_Z\Gamma_Z},\quad
	\Delta(\tilde \nu)  = \frac{i}{p_{\tilde \nu}^2-
			  m^2_{\tilde \nu}},
         \end{equation}
and the longitudinal electron and positron beam polarizations,
$P_{e^-}$ and $P_{e^+}$, respectively,
are included in the coefficients
\begin{eqnarray}
	&&c_L^{ZZ}= L_{e}^2(1-P_{e^-})(1+P_{e^+}), \quad
	c_R^{ZZ}= R_{e}^2(1+P_{e^-})(1-P_{e^+}),\\
	&&c_L^{Z\tilde \nu}=L_{e}(1-P_{e^-})(1+P_{e^+}).
\end{eqnarray}
The contribution (\ref{Zexchange}) from $Z$-exchange is non-zero only
for $\varphi_{\mu}\neq 0,\pi$, whereas the $Z$-$\tilde \nu$ 
interference term, Eq.~(\ref{ZNuinterference}), obtains also
absorptive contributions due to the finite $Z$-width which do not
signal CP violation. These, however, will be eliminated in the 
asymmetry ${\mathcal A}_{\ell}$, Eq.~(\ref{ACP1}).

Analogously to the production matrix,
the chargino decay matrix can be written as
\begin{eqnarray}\label{rhoD}
	\rho_D(\tilde\chi^+_i)_{\lambda_i' \lambda_i} &=&
  \delta_{\lambda_i' \lambda_i} D + \sum_a
       \sigma^{a}_{\lambda_i' \lambda_i}
		\Sigma_D^a.   
\end{eqnarray}
For the chargino decay (\ref{decay_1A}) into an electron or muon sneutrino 
the coefficients are:
\begin{equation}\label{D_1A}
D \; = \; \frac{g^2}{2} |V_{i1}|^2 
	      (m_{\tilde\chi_i^+}^2 -m_{\tilde\nu_{\ell}}^2 ),\quad \quad
	\Sigma^a_{D} \;=\;  \,^{\;\,-}_{(+)} g^2 |V_{i1}|^2 
	m_{\tilde\chi_i^+} (s^a_{\tilde\chi^+_i} \cdot p_{\ell}),
	\quad {\rm for} \;\ell=e,\mu, \\
\end{equation}
where the sign in parenthesis holds for the conjugated process 
$\tilde\chi^-_i \to \ell^-\bar{\tilde\nu}_{\ell}$.
For the decay into the tau sneutrino the coefficients are given by
\begin{equation}\label{D_1B}
D  =  \frac{g^2}{2} (|V_{i1}|^2 +Y_{\tau}^2|U_{i2}|^2)
	(m_{\tilde\chi_i^+}^2 -m_{\tilde\nu_{\tau}}^2 ),\quad 
\Sigma^a_{D} =  \,^{\;\,-}_{(+)} g^2 (|V_{i1}|^2 -Y_{\tau}^2|U_{i2}|^2)
	m_{\tilde\chi_i^+} (s^a_{\tilde\chi^+_i} \cdot p_{\tau}),
\end{equation}
where the sign in parenthesis holds for the conjugated process 
$\tilde\chi^-_i \to \tau^- \bar{\tilde\nu}_{\tau}$.

Inserting the density matrices (\ref{rhoP}) and (\ref{rhoD})
in Eq.~(\ref{amplitude}) leads to:
   \begin{eqnarray} \label{amplitude2}
		|T|^2 &=& 4~|\Delta(\tilde\chi^+_i)|^2~  
			  ( P D + \sum_a  \Sigma_P^a \Sigma_{D}^a ).
				\end{eqnarray}
The cross section and distributions
in the laboratory system are then obtained by integrating 
$|T|^2$ over the Lorentz invariant phase space element, 
\begin{equation}\label{crossection}
d\sigma=\frac{1}{2 s}|T|^2d{\rm Lips} ,
		\end{equation}
where we use the narrow width approximation for the chargino 
propagator.

\section{CP asymmetries
	\label{CP asymmetries}}

Inserting the cross section (\ref{crossection}) in the definition 
of the asymmetry (\ref{AT1}) we obtain: 
\begin{eqnarray}\label{properties}
	{\mathcal A}^{\rm T}_{\ell} 
	 = \frac{\int {\rm Sign}[{\mathcal T_{\ell}}]
		 |T|^2 d{\rm Lips}}
           {\int |T|^2 d{\rm Lips}}
	=  \frac{\int {\rm Sign}[{\mathcal T}_{\ell}]
	\Sigma_P^2 \Sigma_{D}^2 d{\rm Lips}}
          {\int  P D d{\rm Lips}}.
\end{eqnarray}
In the numerator only the CP sensitive contribution 
$\Sigma_P^2 \Sigma_{D}^2$ from chargino polarization perpendicular to 
the production plane  remains, since only
this term contains the triple product
${\mathcal T}_{\ell}=(\vec p_{e^-} \times \vec p_{\tilde\chi^+_i}) 
\cdot \vec p_{\ell}$, Eq.~(\ref{tripleproduct1}).
In the denominator only the term $P D$ remains,
since all spin correlations $\sum_a  \Sigma_P^a \Sigma_{D}^a$ 
vanish due to  the integration over the complete phase space. 
For chargino decay into a tau sneutrino,
$\tilde\chi^+_i \to \tau^+  \tilde\nu_{\tau}$,
the asymmetry $	{\mathcal A}_{\tau}^{\rm T}\propto
	(|V_{i1}|^2 -Y_{\tau}^2|U_{i2}|^2)/
	 (|V_{i1}|^2 +Y_{\tau}^2|U_{i2}|^2)$
is reduced, which follows from the expressions for $D$
and $\Sigma_{D}^2 $, given in  Eq.~(\ref{D_1B}).

With $S_{\ell}$ the standard deviations,
the relative statistical error of the asymmetry
${\mathcal A}_{\ell}^{\rm T}$ is given by 
$\delta {\mathcal A}_{\ell}^{\rm T} = 
\Delta {\mathcal A}_{\ell}^{\rm T}/|{\mathcal A}_{\ell}^{\rm T}| = 
S_{\ell}/(|{\mathcal A}_{\ell}^{\rm T}| \sqrt{N})$ \cite{olaf1},
where $N={\mathcal L} \cdot\sigma$ is the number of events with 
${\mathcal L}$ the integrated luminosity and the cross section 
$\sigma=\sigma_P(e^+e^-\to\tilde\chi^+_i\tilde\chi^-_j) 
\times{\rm BR}(\tilde\chi^+_i \to \ell^+\tilde\nu_{\ell})$.
For the CP asymmetry ${\mathcal A}_{\ell}$,
defined in  Eq.~(\ref{ACP1}), we have
$\Delta {\mathcal A}_{\ell}=\Delta {\mathcal A}_{\ell}^{\rm T}/\sqrt{2} $.
Taking $\delta {\mathcal A}_{\ell}=1$ 
it follows 
$S_{\ell} = |{\mathcal A}_{\ell}| \sqrt{2{\mathcal L}\cdot\sigma}$.
Note that in order to measure ${\mathcal A}_{\ell}$ 
the momentum of $\tilde\chi^+_i$, i.e. the
production plane, has to be determined.
This could be accomplished by measuring the hadronic 
decay of the other chargino $\tilde\chi^-_j$, if the masses
of the charginos and the sneutrinos are known.
Also it is clear that detailed
Monte Carlo studies taking into account background and detector
simulations are necessary to predict the expected accuracy.
However, this is beyond the scope of the present work.

\section{Numerical results
	\label{Numerical results}}

We present numerical results for the  
asymmetries  ${\mathcal A}_{\ell}$ 
for $\ell=e,\mu$, Eq.~(\ref{ACP1}),
and the cross sections 
$\sigma=\sigma_P(e^+e^-\to\tilde\chi^+_1\tilde\chi^-_2 ) \times
{\rm BR}( \tilde\chi^+_1 \to \ell^+\tilde\nu_{\ell})$.
We study the dependence of the asymmetries and cross sections
on the MSSM parameters 
$\mu = |\mu| \, e^{ i\,\varphi_{\mu}}$, 
$M_2$ and $\tan \beta$.
We choose a center of mass energy of   $\sqrt{s} = 800$ GeV
and longitudinally polarized beams with
beam polarizations $(P_{e^-},P_{e^+})=(-0.8,+0.6)$,
which enhance $\tilde\nu_{e}$ exchange in the
production process. This results in larger cross sections
and asymmetries.

We study the decays of the lighter 
chargino $\tilde\chi^+_1$. For the calculation of the chargino 
widths $\Gamma_{\tilde\chi_1^+}$ and the branching ratios 
${\rm BR}( \tilde\chi^+_1 \to\ell^+ \tilde\nu_{\ell})$ 
we include the following two-body decays, 
\begin{eqnarray}
	\tilde\chi^+_1 &\to& 
	W^+\tilde\chi^0_n,~
	e^+\tilde\nu_{e},~
	\mu^+\tilde\nu_{\mu},~
	\tau^+\tilde\nu_{\tau},~
	\tilde e_{L}^+\nu_{e},~
	\tilde\mu_{L}^+\nu_{\mu},~
	\tilde\tau_{1,2}^+\nu_{\tau},
\end{eqnarray}
and neglect three-body decays.
In order to reduce the number of parameters, we assume the 
relation $|M_1|=5/3\,  M_2\tan^2\theta_W $.
For all scenarios we fix the sneutrino and slepton masses, 
$m_{\tilde\nu_{\ell}}=185$~GeV, $\ell =e,\mu,\tau$,
$ m_{\tilde\ell_L}=200$~GeV, $\ell =e,\mu$.
These values are obtained from the renormalization group 
equations \cite{hall}, 
$m_{\tilde\ell_L  }^2 = m_0^2 +0.79 M_2^2
+m_Z^2\cos 2 \beta(-1/2+ \sin^2 \theta_W)$ and
$m_{\tilde\nu_{\ell}}^2 = m_0^2 +0.79 M_2^2 +m_Z^2/2\cos 2 \beta$,
for $M_2=200$~GeV, $m_0=80$~GeV and $\tan\beta=5$.
In the stau sector \cite{thomas} we fix the trilinear scalar coupling
parameter to $A_{\tau}=250$~GeV.
The stau masss are fixed to $m_{\tilde\tau_{1}}=129$~GeV and
$m_{\tilde\tau_{2}}=202$~GeV.

%
\begin{figure}[h]
\setlength{\unitlength}{1cm}
\begin{picture}(10,8)(-0.5,0)
   \put(0,0){\includegraphics{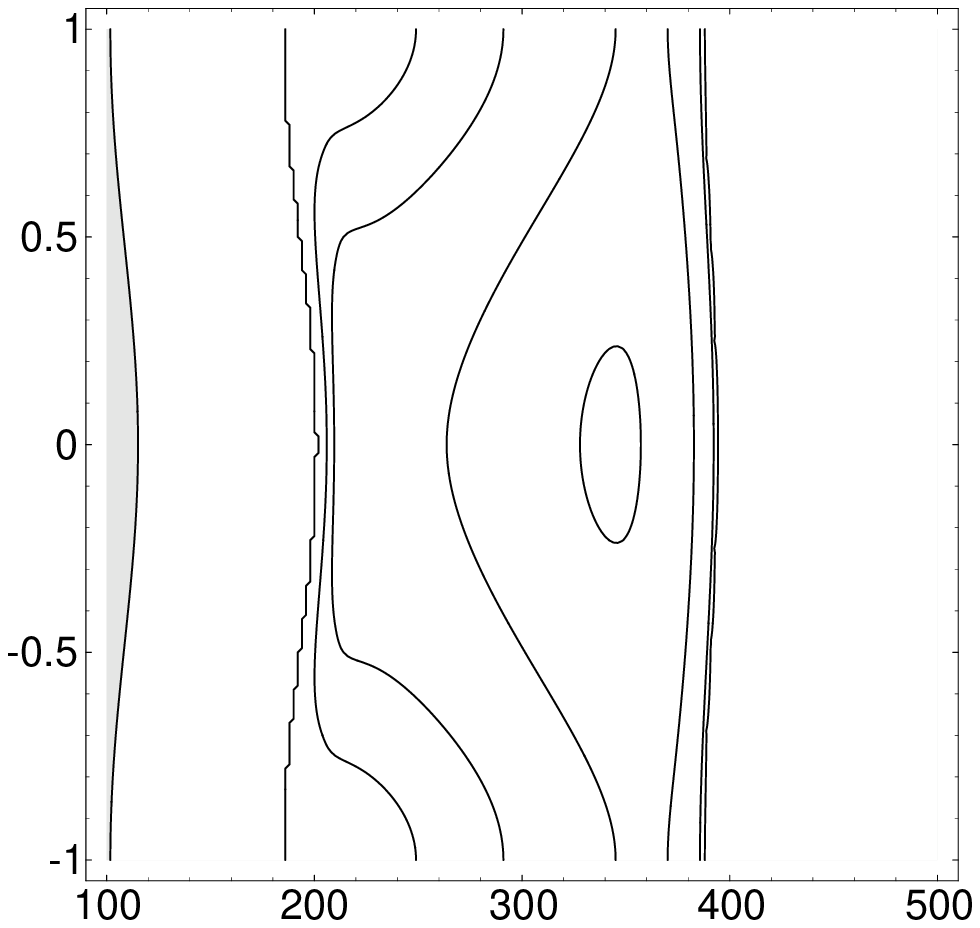}}
	\put(3.5,7.4){\fbox{$\sigma$ in fb}}
	\put(5.5,-0.3){$M_2$~/GeV}
	\put(0,7.4){ $\varphi_{\mu}/\pi$}
	\put(2.25,0.8){\footnotesize 0}
	\put(2.5,1.1){\footnotesize 10}
	\put(2.85,1.65){\footnotesize 20}
	\put(3.4,2.3){\footnotesize 40}
	\put(4.35,3.7){\footnotesize 58}
	\put(1.8,3.7){\begin{picture}(1,1)(0,0)
			\CArc(0,0)(7,0,380)
			\Text(0,0)[c]{{\footnotesize A}}
	\end{picture}}
	\put(6.0,3.7){\begin{picture}(1,1)(0,0)
			\CArc(0,0)(7,0,380)
			\Text(0,0)[c]{{\footnotesize B}}
	\end{picture}}
\put(0.5,-.3){Fig.~\ref{plot_1}a}
	\put(8,0){\includegraphics{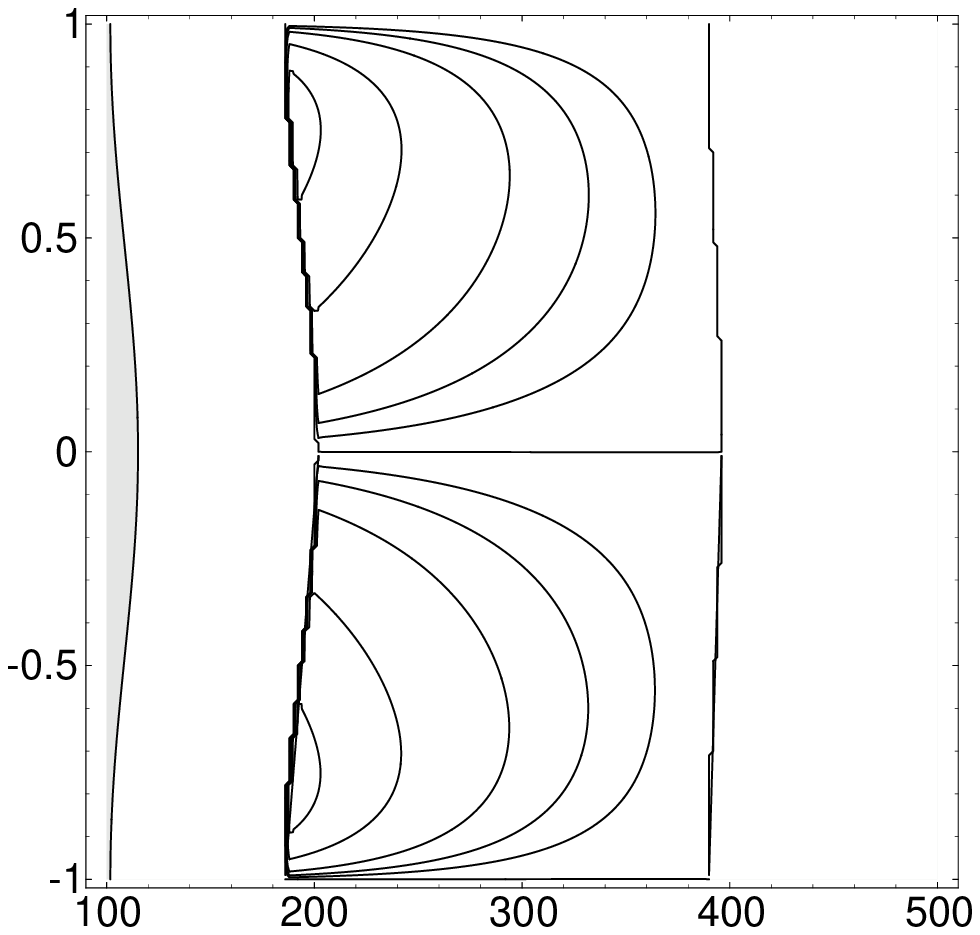}}
	\put(11.,7.4){\fbox{${\mathcal A}_{\ell}$ in \% }}
	\put(13.5,-.3){$M_2$~/GeV}
	\put(8,7.4){ $\varphi_{\mu}/\pi$}
	\put(10.4,1.6){\scriptsize 10}
	\put(11.05,1.8){\footnotesize 5}
	\put(11.75,2.2){\footnotesize 2}
	\put(12.25,2.5){\footnotesize 1}
	\put(12.65,2.8){\footnotesize 0.5}
	\put(13.0,1.1){\footnotesize 0}
	\put(10.45,5.8){\scriptsize -10}
	\put(11.05,5.5){\footnotesize -5}
	\put(11.75,5.15){\footnotesize -2}
	\put(12.3,4.9){\footnotesize -1}
	\put(12.5,4.3){\footnotesize -0.5}
	\put(13.0,6.3){\footnotesize 0}
		\put(9.8,3.7){\begin{picture}(1,1)(0,0)
			\CArc(0,0)(7,0,380)
			\Text(0,0)[c]{{\footnotesize A}}
	\end{picture}}
	\put(14.0,3.7){\begin{picture}(1,1)(0,0)
			\CArc(0,0)(7,0,380)
			\Text(0,0)[c]{{\footnotesize B}}
	\end{picture}}
	\put(8.5,-.3){Fig.~\ref{plot_1}b}
\end{picture}
\vspace*{.5cm}
\caption{
	Contour lines of 
	$\sigma=\sigma_P(e^+e^-\to\tilde\chi^+_1\tilde\chi^-_2) 
	\times {\rm BR}( \tilde\chi^+_1 \to \ell^+\tilde\nu_{\ell})$, 
	summed over $\ell = e,\mu$,
	(\ref{plot_1}a), and the asymmetry ${\mathcal A}_{\ell}$ 
	for $\ell =e$ or $\mu$ (\ref{plot_1}b),
	in the $M_2$--$\varphi_{\mu}$ plane for 
	$|\mu|=400$~GeV, $\tan\beta =5$, 
	$m_{\tilde\nu_{\ell}}=185$~GeV, 
	$\sqrt{s}=800$ GeV and $(P_{e^-},P_{e^+})=(-0.8,0.6)$.
	The gray  area is excluded by $m_{\tilde\chi_1^{\pm}}<104 $ GeV.
	The area A is kinematically forbidden by
	$m_{\tilde\nu_{\ell}}+m_{\tilde\chi^0_1}> m_{\tilde\chi^+_1}$.
	The area B is kinematically forbidden by
	$m_{\tilde\chi^+_1}+m_{\tilde\chi^-_2}>\sqrt{s}$.
	\label{plot_1}}
\end{figure}
In Fig.~\ref{plot_1}a we show the contour lines of the 
cross section for chargino production and decay
$\sigma=\sigma_P(e^+e^-\to\tilde\chi^+_1\tilde\chi^-_2) 
	\times {\rm BR}( \tilde\chi^+_1 \to \ell^+\tilde\nu_{\ell})$
in the $M_2$--$\varphi_{\mu}$ plane
for $|\mu|=400$ GeV and  $\tan\beta =5$. 
The production cross section 
$\sigma_P(e^+e^-\to\tilde\chi^+_1\tilde\chi^-_2)$
can attain values from $10$ fb to $150$ fb
and ${\rm BR}( \tilde\chi^+_1 \to \ell^+\tilde\nu_{\ell})$,
summed over $\ell = e,\mu$, can be as large as 50\%. Note that $\sigma$ 
is very sensitive to $\varphi_{\mu}$, which has been exploited 
in \cite{choi1,choigaiss} to constrain $\cos(\varphi_{\mu})$.

The $M_2$--$\varphi_{\mu} $ dependence of the CP asymmetry
${\mathcal A}_{\ell}$ for $\ell =e$ or $\mu$ is shown in Fig.~\ref{plot_1}b.
The asymmetry can be as large as 10\% and it does, however, not attain 
maximal values for $\varphi_{\mu}=0.5\pi$, which one would naively expect.
The reason is that  ${\mathcal A}_{\ell}$ is proportional to a product of
a CP odd ($\Sigma_P^2$) and a CP even factor ($\Sigma_{D}^2$),
see Eq.~(\ref{properties}). The  CP odd (CP even) factor has
as sine-like (cosine-like) dependence on  $\varphi_{\mu}$.
Thus the maximum of ${\mathcal A}_{\ell}$ is shifted  
towards $\varphi_{\mu}=\pm\pi$ in Fig.~\ref{plot_1}b. Phases close  to the
CP conserving points, $\varphi_{\mu}= 0,\pm \pi$, are favored by
the experimental upper limits on the EDMs.
For example in the constrained MSSM, we have
$|\varphi_{\mu}|\lsim \pi/10$ \cite{edms}.
However, the restrictions are very model dependent, e.g., 
if also lepton flavor violating terms are included \cite{BMPW},
the restrictions may disappear. In order to show the full phase dependence 
of the asymmetries, we have relaxed the EDM restrictions for this purpose. 

%
\begin{figure}[h]
\setlength{\unitlength}{1cm}
\begin{picture}(10,8)(-0.5,0)
   \put(0,0){\includegraphics{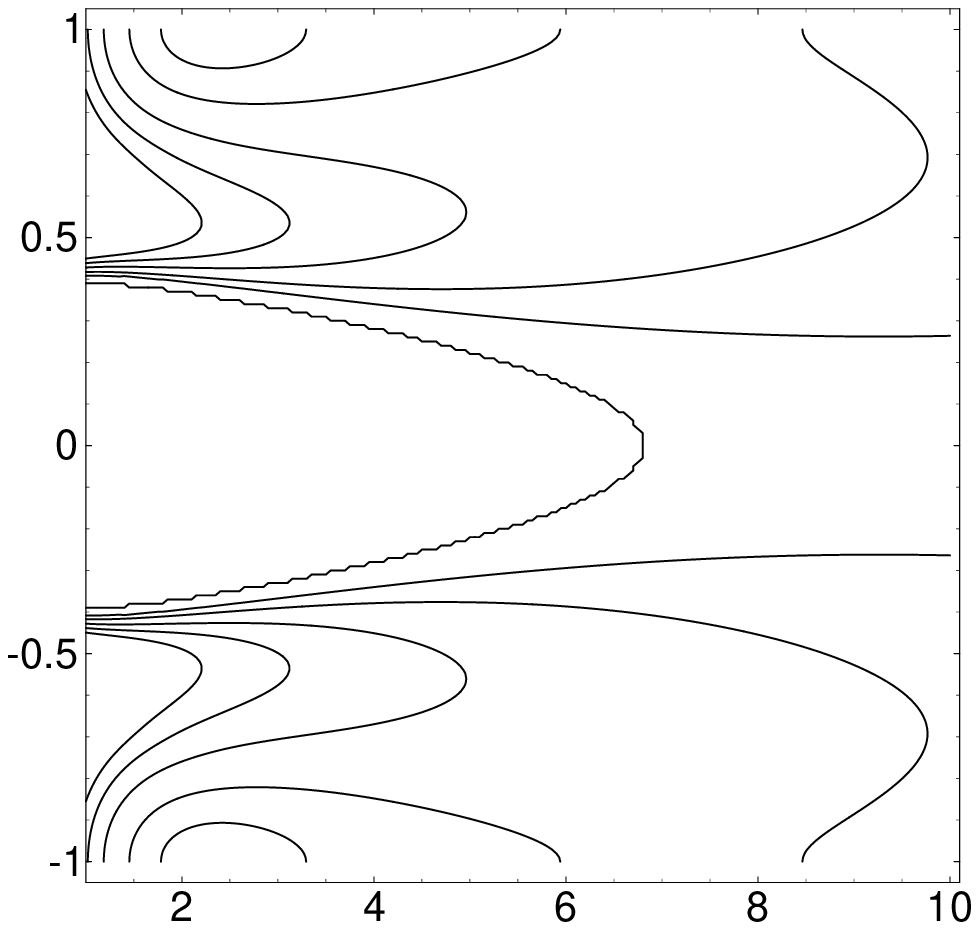}}
	\put(3.5,7.4){\fbox{$\sigma$ in fb}}
	\put(6.3,-0.3){$\tan\beta$}
	\put(0,7.4){ $\varphi_{\mu}/\pi$}
	\put(1.1,2.0){\footnotesize 20}
	\put(1.75,2.05){\footnotesize 15}
	\put(3.0,2.0){\footnotesize 10}
	\put(5.6,2.1){\footnotesize 5}
	\put(6.3,3.1){\footnotesize 2}
	\put(1.8,0.8){\footnotesize 2}
	\put(3.1,0.9){\footnotesize 5}
	\put(4.8,3.7){\footnotesize 0}
	\put(1.1,5.3){\footnotesize 20}
	\put(1.7,5.3){\footnotesize 15}
	\put(3.0,5.35){\footnotesize 10}
	\put(5.6,5.2){\footnotesize 5}
	\put(6.3,4.2){\footnotesize 2}
	\put(1.8,6.5){\footnotesize 2}
	\put(3.1,6.4){\footnotesize 5}
		\put(2.0,3.7){\begin{picture}(1,1)(0,0)
			\CArc(0,0)(7,0,380)
			\Text(0,0)[c]{{\footnotesize A}}
	\end{picture}}
\put(0.5,-.3){Fig.~\ref{plot_2}a}
	\put(8,0){\includegraphics{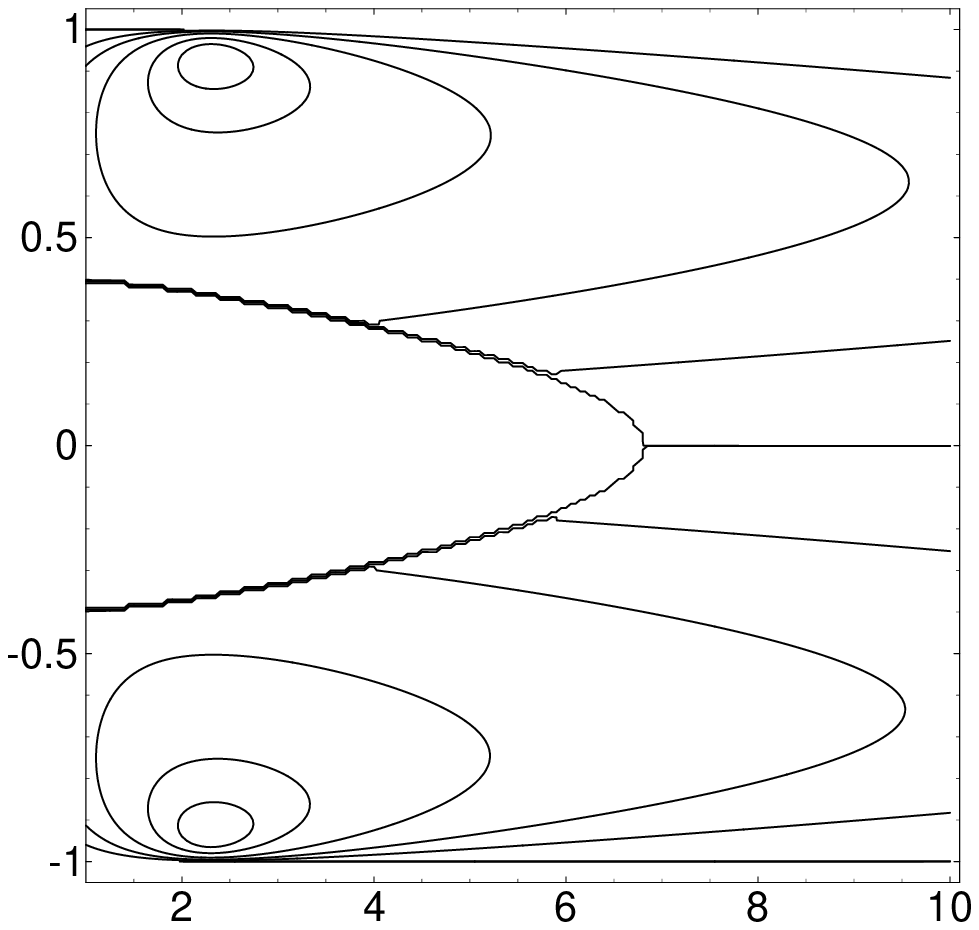}}
	\put(11.,7.4){\fbox{${\mathcal A}_{\ell}$ in \% }}
	\put(14.3,-.3){$\tan\beta$}
	\put(8,7.4){ $\varphi_{\mu}/\pi$}
	\put(9.55,1.){\scriptsize 30}
	\put(10.3,1.4){\scriptsize 20}
	\put(11.4,1.9){\footnotesize 10}
	\put(14.2,1.8){\footnotesize 5}
	\put(14.3,2.75){\footnotesize 2.5}
	\put(14.3,1.2){\footnotesize 2.5}
	\put(14.4,3.85){\footnotesize 0}
	\put(14.65,0.85){\footnotesize 0}
	\put(9.5,6.4){\scriptsize -30}
	\put(10.3,6.0){\scriptsize -20}
	\put(11.5,5.6){\footnotesize -10}
	\put(14.2,5.55){\footnotesize -5}
	\put(14.2,4.6){\footnotesize -2.5}
	\put(14.3,6.1){\footnotesize -2.5}
	\put(10.,3.7){\begin{picture}(1,1)(0,0)
			\CArc(0,0)(7,0,380)
			\Text(0,0)[c]{{\footnotesize A}}
	\end{picture}}
	\put(8.5,-.3){Fig.~\ref{plot_2}b}
\end{picture}
\vspace*{.5cm}
\caption{
	Contour lines of 
	$\sigma=\sigma_P(e^+e^-\to\tilde\chi^+_1\tilde\chi^-_2) 
	\times {\rm BR}( \tilde\chi^+_1 \to \ell^+\tilde\nu_{\ell})$, 
	summed over $\ell = e,\mu$, (\ref{plot_2}a), 
	and the asymmetry ${\mathcal A}_{\ell}$ for 
	$\ell =e$ or $\mu$ (\ref{plot_2}b),
	in the $\tan\beta$--$\varphi_{\mu}$ plane for 
	$M_2=200$~GeV, $|\mu|=400$~GeV, $m_{\tilde\nu_{\ell}}=185$~GeV,
	$\sqrt{s}=800$ GeV and $(P_{e^-},P_{e^+})=(-0.8,0.6)$.
	The area A  is kinematically forbidden by
	$m_{\tilde\nu_{\ell}}+m_{\tilde\chi^0_1}> m_{\tilde\chi^+_1}$.
	\label{plot_2}}
\end{figure}
For $M_2=200$ GeV, we show the 
$\tan\beta $--$\varphi_{\mu} $ dependence of 
$\sigma$ and ${\mathcal A}_{\ell}$ in Figs.~\ref{plot_2}a,b.
%
The asymmetry can reach values up to 30\% and  shows a
strong  $\tan\beta$ dependence and decreases with increasing $\tan\beta$. 
The feasibility of measuring the asymmetry depends also on the cross 
section $\sigma=\sigma_P(e^+e^-\to\tilde\chi^+_1\tilde\chi^-_2)\times
 {\rm BR}(\tilde\chi^+_1 \to \ell^+\tilde\nu_{\ell})$,
Fig.~\ref{plot_2}a, which attains values up to 20 fb.

%
\begin{figure}[h]
\setlength{\unitlength}{1cm}
\begin{picture}(10,8)(-0.5,0)
   \put(0,0){\includegraphics{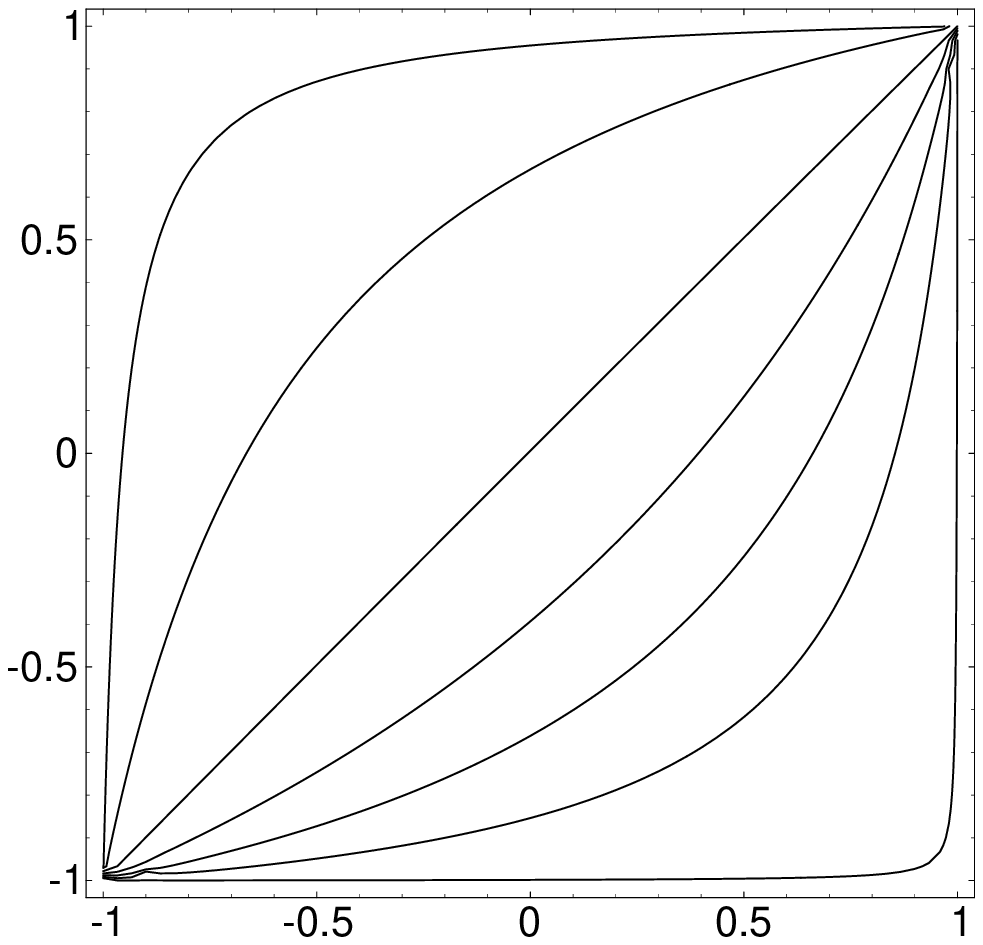}}
	\put(3.5,7.4){\fbox{${\mathcal A}_{\ell}$ in \% }}
	\put(6.5,-0.3){$ P_{e^-}$}
	\put(0.5,7.4){$ P_{e^+} $ }
	\put(1.3,6.2){\footnotesize -7.3}
	\put(2.6,5.1){\footnotesize -7}
	\put(3.6,3.8){\footnotesize -6}
	\put(4.3,3.2){\footnotesize -5}
	\put(4.85,2.7){\footnotesize -4}
	\put(5.45,2.2){\footnotesize -3}
	\put(6.5,1.0){\footnotesize -2}
	
	\put(.,.){\footnotesize }
\put(0.5,-.3){Fig.~\ref{plot_3}a}
	\put(8,0){\includegraphics{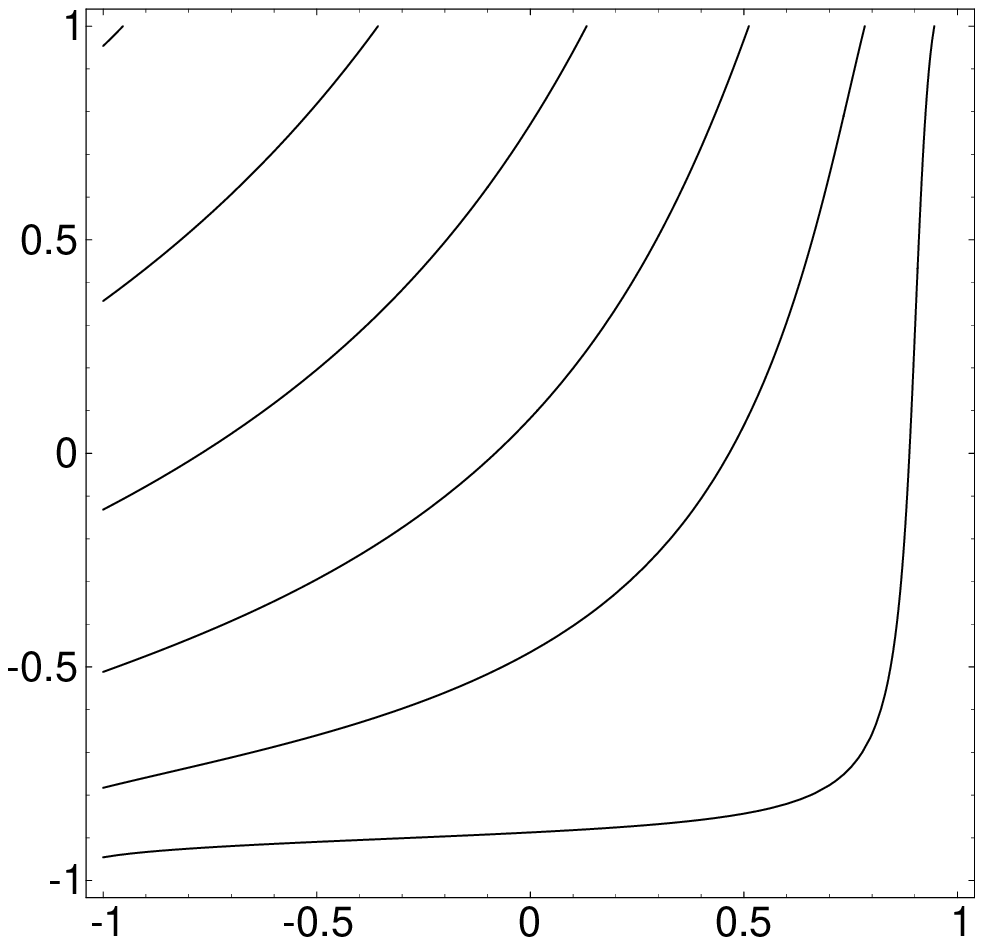}}
	\put(10.3,7.4){\fbox{$S_{\ell} =|{\mathcal A}_{\ell}| 
			\sqrt{2{\mathcal L}\cdot\sigma}$  }}
	\put(14.5,-.3){$ P_{e^-}$ }
	\put(8.5,7.4){$  P_{e^+}$ }
	\put(9.,6.5){\footnotesize 6}
	\put(9.7,5.7){\footnotesize 5}
	\put(10.7,4.8){\footnotesize 4}
	\put(11.7,4.0){\footnotesize 3}
	\put(12.7,3.1){\footnotesize 2}
	\put(14.,1.6){\footnotesize 1}
\put(8.5,-.3){Fig.~\ref{plot_3}b}
\end{picture}
\vspace*{.5cm}
\caption{
	Contour lines of  
	the asymmetry ${\mathcal A}_{\ell}$ for $\ell =e$ or $\mu$
	(\ref{plot_3}a),
	and the standard deviations $S_{\ell}$ (\ref{plot_3}b),
	for $e^+e^-\to\tilde\chi^+_1\tilde\chi^-_2;~ 
	\tilde\chi^+_1 \to \ell^+\tilde\nu_{\ell}$
	in the $ P_{e^-}$--$P_{e^+}$ plane for $\varphi_{\mu}=0.9\pi$,
	taking $|\mu|=400$~GeV, $M_2=200$~GeV,
	$\tan \beta=5$, $m_{\tilde\nu_{\ell}}=185$~GeV, 
	$\sqrt{s}=800$ GeV and ${\mathcal L}=500~{\rm fb}^{-1}$.
	\label{plot_3}}
\end{figure}
For the phase $\varphi_{\mu}=0.9\pi$ and $\tan \beta=5$,
we study the beam polarization dependence of ${\mathcal A}_{\ell}$,
which can be strong as shown in Fig.~\ref{plot_3}a. 
An electron beam polarization $P_{e^-}>0$ and a positron
beam polarization $P_{e^+}<0$ enhance the channels with $\tilde\nu_{e}$
exchange in the chargino production process. 
For e.g. $(P_{e^-},P_{e^+})=(-0.8,0.6)$
the asymmetry can attain -7\%, Fig.~\ref{plot_3}a,
with  $\sigma_P(e^+e^-\to\tilde\chi^+_1\tilde\chi^-_2)\approx10$ fb
and ${\rm BR}(\tilde\chi^+_1 \to \ell^+\tilde\nu_{\ell})\approx50\%$,
summed over $\ell = e,\mu$.
The cross section $\sigma=\sigma_P(e^+e^-\to\tilde\chi^+_1\tilde\chi^-_2)\times
 {\rm BR}(\tilde\chi^+_1 \to \ell^+\tilde\nu_{\ell})$
 ranges between 2.3 fb for $(P_{e^-},P_{e^+})=(0,0)$ and
6.8 fb for $(P_{e^-},P_{e^+})=(-1,1)$.
The standard deviations of ${\mathcal A}_{\ell}$,  given by
$S_{\ell} =|{\mathcal A}_{\ell}| \sqrt{2{\mathcal L}\cdot\sigma}$,
are shown in  Fig.~\ref{plot_3}b for ${\mathcal L}=500~{\rm fb}^{-1}$.
We have $S_{\ell}\approx 5$ for $(P_{e^-},P_{e^+})=(-0.8,0.6)$,
and thus ${\mathcal A}_{\ell}$ could be accessible at a linear
collider, even for $\varphi_{\mu}=0.9\pi$, by using polarized beams.

\section{Summary and conclusions
	\label{Summary and conclusion}}

We have studied CP violation in chargino production with longitudinally 
polarized beams, $e^+e^- \to\tilde\chi^+_i  \tilde\chi^-_j$,
and subsequent two-body decay  of one  chargino
into the sneutrino $\tilde\chi^+_i \to \ell^+\tilde\nu_{\ell}$.
We have defined the T odd asymmetries 
$ {\mathcal A}_{\ell}^{\rm T}$ of the triple product 
$(\vec p_{e^-} \times \vec p_{\tilde\chi^+_i}) \cdot \vec p_{\ell}$.
The CP odd asymmetries 
${\mathcal A}_{\ell} = \frac{1}{2}({\mathcal A}_{\ell}^{\rm T}-
	 \bar{\mathcal A}_{\ell}^{\rm T})$,
where $\bar{\mathcal A}_{\ell}^{\rm T}$ denote the CP conjugated
of ${\mathcal A}_{\ell}^{\rm T}$, are sensitive to the phase  
$\varphi_{\mu}$ of the Higgsino mass parameter $\mu$. 
At tree level, the asymmetries have 
large CP sensitive contributions from spin correlation effects in the 
production of an unequal pair of charginos.
In a numerical discussion for 
$e^+e^- \to\tilde\chi^+_1  \tilde\chi^-_2$ production, we have found
that  ${\mathcal A}_{\ell}$ for $\ell =e$ or $\mu$  can attain values 
up to 30\%. By analyzing the statistical errors, we have shown that,
even for of e.g. $\varphi_{\mu}\approx0.9 \pi$, the 
asymmetries could be accessible in future  $e^+e^-$ collider 
experiments in the 800 GeV range with high luminosity and 
longitudinally polarized beams.

\section{Acknowledgments}

This work was supported by the 'Deutsche Forschungsgemeinschaft'
(DFG) under contract Fr 1064/5-2.
This work was also supported by the `Fonds zur
F\"orderung der wissenschaftlichen Forschung' (FWF) of Austria, project
No. P16592-N02, and by the European Community's
Human Potential Programme under contract HPRN-CT-2000-00149.

\end{document}